\documentclass[fleqn,10pt]{wlscirep}

\usepackage{graphicx} 
\usepackage{subfigure}
\usepackage{tikz}
\usepackage{bm}
\usepackage{amsmath}
\usepackage{amsfonts}
\usepackage{mathrsfs}

\title{Do ultrafast exciton-polaron decoherence dynamics govern photocarrier generation efficiencies in polymer solar cells?  }
\author[1,+]{Eleonora~Vella} 
\author[2,+]{Hao~Li}  
\author[1]{Pascal~Gr{\'e}goire} 
\author[3]{Sachetan~M.~Tuladhar}  
\author[3]{Michelle~S.~Vezie}  
\author[3]{Sheridan~Few}  
\author[1]{Claudia~M.~Baz{\'a}n}  
\author[3]{Jenny~Nelson}
\author[1,3,*]{Carlos~Silva-Acu\~na}  
\author[2,1,*]{Eric~R.~Bittner}  

\affil[1]{Department of Physics and Regroupement qu{\'e}b{\'e}cois sur les mat{\'e}riaux de pointe, Universit{\'e} de Montr{\'e}al, C.P. 6128, Succursale centre-ville, Montr{\'e}al  H3C~3J7, Canada}
\affil[2]{Department of Chemistry, University of Houston, Houston, Texas 77204, USA}
\affil[3]{Department of Physics, Blackett Laboratory, Imperial College London, London SW7~2AZ, United Kingdom}
\affil[*]{to whom correspondence should be addressed}

\begin{abstract}
All-organic-based photovoltaic solar cells have attracted considerable attention because of their low-cost processing and short energy payback time.
In such systems the primary dissociation of an optical excitation into a pair of photocarriers has  been recently shown to be extremely rapid and efficient,  
but the physical reason for this remains unclear. Here, two-dimensional photocurrent excitation spectroscopy, a novel non-linear optical spectroscopy, is used to probe the ultrafast coherent decay of photoexcitations into charge-producing states in a polymer:fullerene based solar cell. The two-dimensional photocurrent spectra are interpreted by introducing a theoretical model for the description of the coupling of the electronic states of the system to an external environment and to the applied laser fields. The experimental data show no cross-peaks in the two-dimensional photocurrent spectra, as predicted by the model for coherence times between the exciton and the photocurrent producing states of 20\,fs or less.
\end{abstract}
\begin{document}

\flushbottom
\maketitle
%
%
%
\thispagestyle{empty}

\section*{Introduction}

Advances in the fabrication of organic polymer-based 
photovoltaic (OPV) cells have led to power conversion efficiencies
(PCE) in excess of 10\% in certain polymer:fullerene devices,
making them viable materials for light-harvesting devices.\cite{Bredas31012014}
However, 
the fundamental photophysical pathways that connect
the absorption of light to the production of charge carriers
 remain elusive in spite of rigorous experimental and theoretical 
investigation. 
OPVs based on the so-called bulk heterojunction consist of
blends of electron donor (usually a $\pi$-conjugated polymer) 
and electron acceptor (typically a fullerene derivative or a second polymer) materials.

The bottleneck that has attracted the attention of theorists and 
experimentalists alike is that the primary photoexcitations
are tightly bound intramolecular excitons with 
high-binding energy ($\approx 500$\,meV). 
Subsequently, the lowest-energy intermolecular charge-transfer (CT) states that
occur at the phase-boundary between donor and acceptor materials
also have binding energies in the range of 300--400\,meV~\cite{Gelinas:2011vn}.
The general paradigm for some time has been that the CT states
are the primary precursors to photocarriers.
However, in order to produce photocurrent, the electrons and holes produced 
by photoexcitation must separate
far enough such that their binding energy could be small as compared to the thermal energy near ambient conditions.

Recent spectroscopic measurements on organic photovoltaic 
systems report that charged photoexcitations 
can be generated on $\leq 100$-fs timescales~\cite{Sariciftci:1994kx,
Banerji:2010vn,
Tong2010,
Sheng2012,
Jailaubekov:2013fk,
Grancini:2013uq,
doi:10.1021/jz4010569,
doi:10.1021/jp4071086,
Banerji:2013ej}. 
Delocalization of the transferred electron 
amongst several fullerenes diminishes its Coulombic attraction to a hole localized in 
the polymer (donor) phase. Transient absorption experiments 
by G\'elinas {\em et~al.}, in which Stark-effect signatures in transient absorption
 spectra were analysed to probe the local electric field as  charge separation proceeds, 
indicate that electrons and holes separate by $\sim40$\AA\ 
over the first 100\,fs\ and evolve further on picosecond 
timescales to produce unbound charge pairs~\cite{Gelinas:2013fk}. 
Model calculations based upon  Fermi's golden rule also indicate 
that such rapid time-scales are only consistent with a model 
of a highly-delocalised charge in the acceptor phase~\cite{Gelinas:2013fk}. 
Concurrently, Provencher {\em et~al.} demonstrated, via femtosecond stimluated Raman spectroscopic measurements, the emergence of clear polaronic vibrational signatures on sub-100-fs on the polymer backbone, with very limited molecular reorganization or vibrational relaxation following the ultrafast step~\cite{Provencher:fk}. Such spectacularly and apparently universally rapid through-space charge transfer between excitons on the polymer backbone and acceptors across the heterojunction would be difficult to rationalize within Marcus theory using a localised basis without invoking unphysical distance dependence of tunnelling rate constants~\cite{Barbara:1996uc}. 
In addition, intramolecular electron-vibration interactions in organic 
conjugated polymers are known to be relatively strong 
and do contribute to the photophysical dynamics.
Recent observations of quantum beating with a time-period of 23\,fs in a prototypical 
polymer:fullerene OPV system suggest that C=C stretching 
modes of the polymer and the pinching mode of the fullerene 
are as important as the off-diagonal electronic transfer integral itself\cite{Falke30052014}, a scenario not unlike what occurs in photosynthetic light-harvesting systems.\cite{Ishizaki:2009gd}
Simlarly, Song~et~al.\ demonstrated, by means of two-dimensional coherent spectroscopy, the role that such vibrational coherence plays in  ultrafast charge generation processes in polymer:fullerene systems.~\cite{Song:1uq} 
Such observations
underscore the importance of coherent vibronic coupling between electronic and nuclear degrees of freedom in charge delocalization and transfer in a noncovalently bound reference system.\cite{Rozzi:2013fk}

We recently proposed that quantum mechanical tunnelling processes brought
about by environmental fluctuations couple photoexcitations {\em directly} to photocurrent  producing charge-transfer states 
on a sub 100-fs timescales. \cite{Bittner:2014aa}
Our heuristic model is supported by a recent 
numerically exact study of a model spin-boson system with separated 
diagonal and off-diagonal baths by Yao {\em et al.}\cite{Yao:2015aa}
This work indicates that there exists a critical parametric regime that produces 
neither localized nor delocalized states
and is free of quantum decoherence, suggesting that the existence
 of such decoherence-free subspaces is critical to the formation of quantum coherent exchange between the exciton and the CS states.\cite{Zhao-2014,Yao:2015aa}

In the present work, we exploit the superior detection sensitivity of ultrafast photocurrent probes\:\cite{Bakulin:2016rt} to explore photocarrier generation dynamics by means of two-dimensional (2D) photocurrent excitation spectroscopy (2DPCE). This is a novel nonlinear optical spectroscopy in which the two-dimensional excitation correlation spectrum is measured as a function of time after initial excitation with a femtosecond pulse sequence via photocurrent.\cite{Nardin:2013bf,Karki:2014eu} In general, 2D electronic spectroscopies permit identification of homogeneous and inhomogeneous spectral lineshape contributions, and because they are nonlinear optical techniques, they can reveal the existence of couplings between distinct excited states through the presence of cross peaks (off-diagonal signals) in the 2D correlation spectrum. Because of these unique advantages over linear techniques, 2DPCE is extremely valuable in the investigation of the photocharge generation dynamics in materials in which photocarriers are not produced directly by inter-band transitions as in bulk semiconductors, but in which precursors to photocarrier pairs, such as excitons in molecular semiconductors, are the primary photoexcitations. We use 2DPCE in order to explore the ultrafast decoherence dynamics between mixed exciton and delocalised polaron states that we proposed to be photocarrier precursors in ref.~\citenum{Bittner:2014aa}. 
%
We begin with a brief overview of a theoretical model describing
salient electronic states of this system and their coupling to
an external environment and to the applied laser fields.  This
will serve as an important touchstone in interpreting our 
experimental data.   We then describe the results of our 
2DPCE experiments on a working organic polymer based solar cell
based upon a polymer:fullerene blend.  

\section{Results}
\subsection*{Theoretical model}
\begin{figure}
\includegraphics[width=\textwidth]{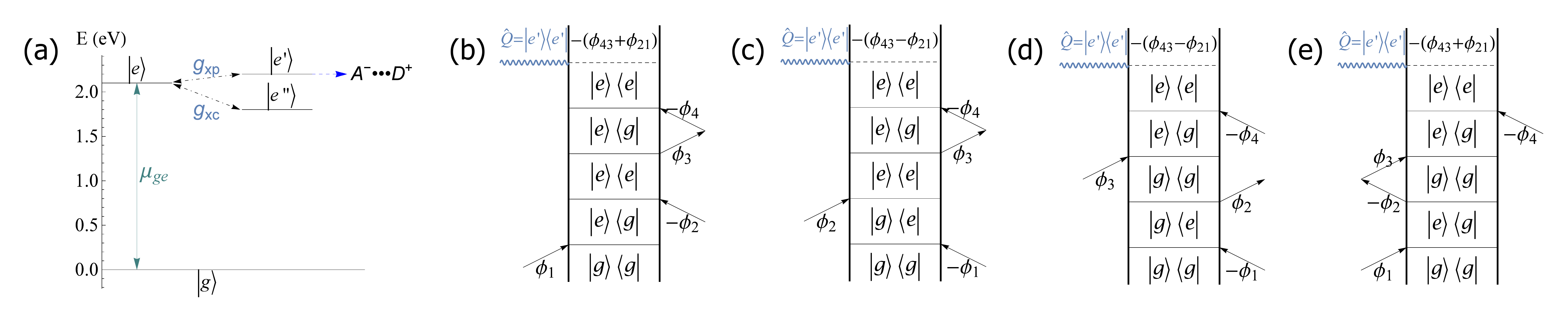}
\caption{(a) Schematic of the relevant energy levels in our model along with their couplings.
(b-e) Double-sided Feynman diagrams contributing to the 4th order 
population.  Incoming and outgoing arrows indicate interactions with the laser fields
along with their phase. $\hat{Q}$ is the final projection onto the outgoing polaron states that 
contribute to the photocurrent. 
}
\label{Energy_Feynman_diagrams}
\end{figure}

We begin by developing a model for the salient 
electronic states of an organic heterojunction system as 
sketched in Fig.~\ref{Energy_Feynman_diagrams}a, in which 
we assume that the primary photoexcitation (exciton) 
is produced by the absorption of a photon from the electronic 
ground state of the system and can decay into 
either a charge-transfer (CT) state
pinned to the heterojunction interface, or into delocalised charge-separated
(CS), or polaron pair, states.\cite{Bittner:2014aa} We also assume that population in 
the CS states is ultimately responsible for any photocurrent 
produced by the actual device.  
Consequently, in the limit of no
other channels for carrier loss, photocurrent is at least 
proportional to the net population transferred to the CS states.  
Moreover, both spectroscopic measurements and quantum chemical studies place 
the CT state energetically well below  the threshold for polaron formation 
such that these CT states cannot decay into 
CS states via thermally activated processes.\cite{Gelinas:2011vn} 

\begin{figure}[t]
\centering
\includegraphics[width=0.75\textwidth]{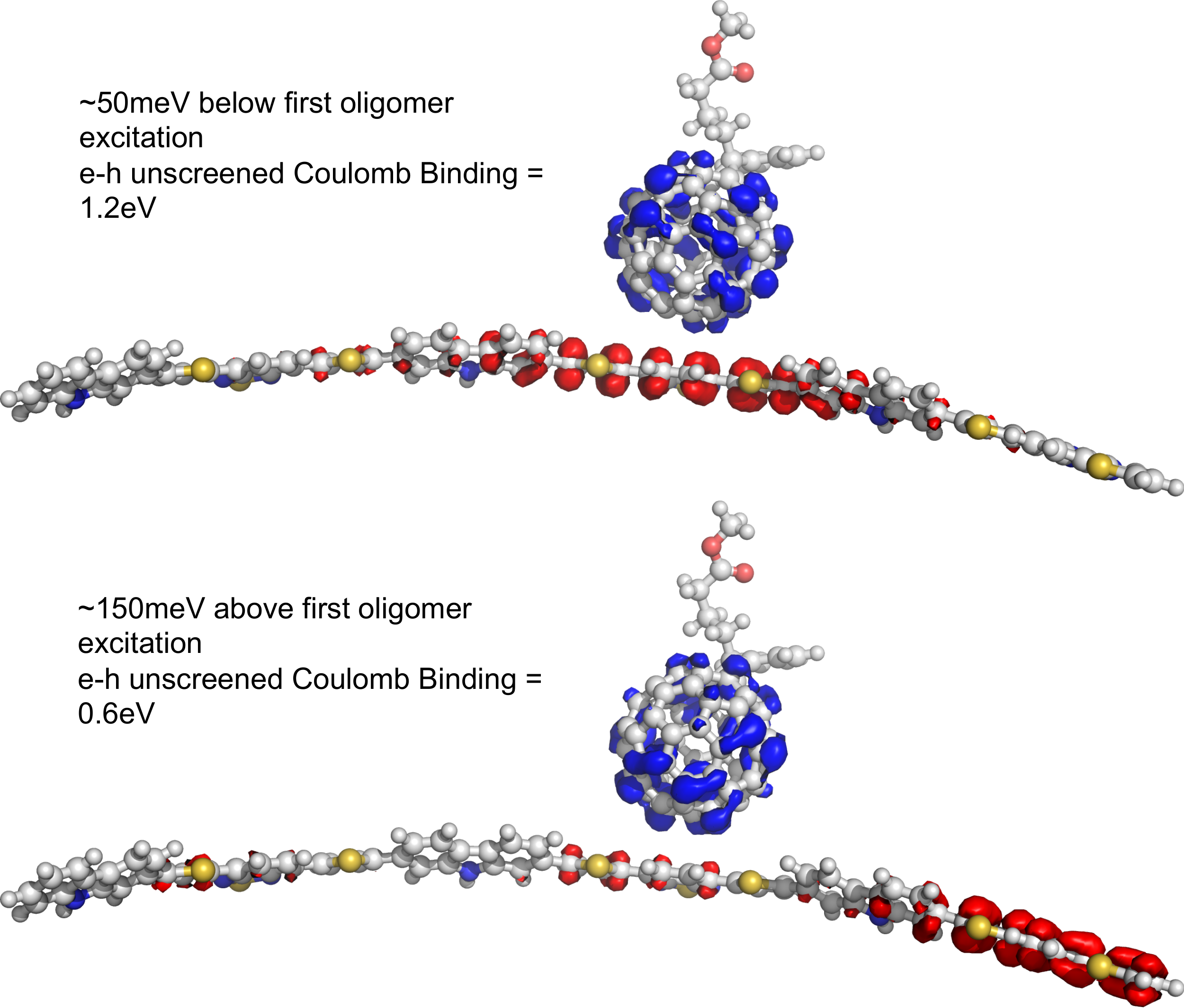}
\caption{Redistribution of charge, excitation energy, and unscreened Coulomb binding energy for two excitations of a CDTBT oligomer:PCBM molecule pair. Isosurfaces show change in charge distribution between ground state and excited states (electron density moves from red to blue regions).}
\label{fig_isosurfaces}
\end{figure}
To model any particular physical system, we need information about the energy and nature of the excited states at the donor:acceptor interface where charge separation occurs. The challenge of calculating the properties of such states is complicated by the wide variety of interfacial molecular packing arrangements and by the size of the interfacial region. This remains an active area of research.\cite{Few:2015fk} In the present case, we use the simplest possible model of the donor: acceptor interface, a complex consisting of an oligomer of the donor polymer and a single fullerene derivative, and calculate the excited states in vacuum using time-dependent density functional theory (TDDFT) with B3LYP/6-31g*. We classify the resulting states as excitonic, charge-transfer (CT) or mixed according to what fraction of an electron charge lies on the fullerene, and we study the dependence of the energy and type of these states on the relative position of oligomer and fullerene molecule. Although the method is relatively simple, it was shown previously 
by Few~et~al.\ to reproduce the observed trend in lowest excitation energies (determined through electroluminescence) with the chemical structure of donor and acceptor constituent materials \cite{Few:2014fk}. Our calculation method is described in detail in ref.~\citenum{Few:2014fk}. 

Here, we focus on the donor: acceptor system that we study experimentally, the polymer poly(N-9''-hepta-decanyl-2,7-carbazole-alt-5,5-(4',7'-di-2-thienyl-2',1',3'-benzothiadiazole)) (PCDTBT) combined with the fullerene derivative phenyl C-61 butyric acid methyl ester (PCBM). 
We model the polymer as a trimer in its optimised geometry and study the excited state spectrum for different relative positions of fullerene and oligomer. For each fullerene alignment, we calculate three CT states (more then 0.9 e transferred from oligomer to fullerene) at excitation energies below that of the first oligomer singlet, and a number of ``mixed  states" (between 0.1 e and 0.9 e transferred from oligomer to fullerene) at energies close to that of oligomer singlet transitions. When the PCBM molecule is aligned with the benzothiadiazole (BT) unit of the oligomer, we calculate an additional CT state with an excitation energy $\sim 150$\,meV above that of the oligomer singlet. For each CT state we calculate the Coulomb interaction due to partial charges on each molecule, $U_{B}$, which gives an indication of the energy required to generate a free-charge pair. (Note that dielectric screening is not taken into account when calculating this value, and it should be used as a guide to relative binding of states, rather than as an absolute value.) As illustrated in Fig.~\ref{fig_isosurfaces}, in the case when the PCBM molecule aligns with the BT unit of the oligomer we find the higher energy CT state to have a lower $U_{B}$ (0.6\,eV) than lower energy CT states ($\sim$ 1.2\,eV). A similar trend in decreasing Coulombic binding energy with increasing CT state energy has been reported in a variety of other donor: acceptor blends \cite{bakulin2012}, \cite{Few:2014fk}.

\begin{figure}
\centering
\includegraphics[width=0.75\textwidth]{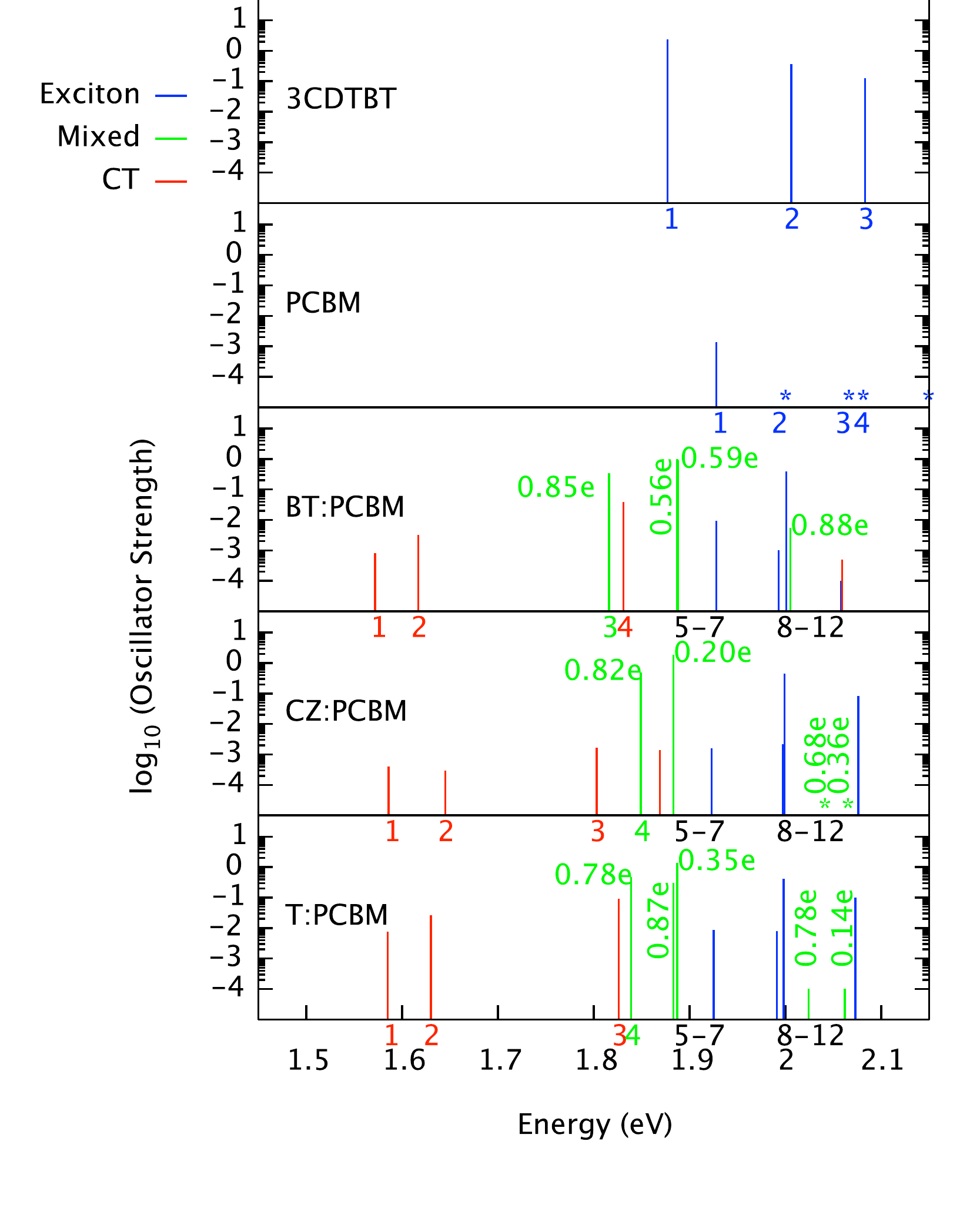}
\caption{Calculated excitation spectrum for a 3CDTBT oligomer and a PCBM molecule, each alone, and as a molecule pair. Excitations are coloured according to degree of charge transfer exhibited. 
Single molecule excitations (CT $\leq$ 0.10 e) are blue, complete charge transfer excitations (CT $\geq$ 0.90 e) are red, and mixed excitations (0.10 e \textless  CT \textless  0.90 e) are green.}
\label{fig_3CDTBT_PCBM_spectrum}
\end{figure}

Fig.~\ref{fig_3CDTBT_PCBM_spectrum} shows the spectrum of excited states for the donor trimer (3CDTBT), the PCBM molecule, and the 3CDTBT:PCBM molecule pair with the fullerene located above the benzothiadiazole unit (BT), the carbazole unit (CZ) and the thiophene unit (T), calculated as described in ref.~\citenum{Few:2014fk}.

To cast our quantum-dynamical model involving excitations in Fig.~\ref{Energy_Feynman_diagrams}a into a succinct theoretical framework, 
we consider the time-evolution of the electronic states of the
model system as an open quantum system under influence of a series of collinear phase-modulated laser pulses.  
In our calculations, all quantum operators $A(t)$ are
propagated according to the Lindblad master equation
\begin{align}
 \label{eqn:LindbladEq}
 \frac{d A}{d t}=\frac{i}{\hbar}\left[H_S,A\right]+{\mathscr L}_D(A),
\end{align}
in which $H_S$ is the unperturbed Hamiltonian of the electronic system in its second quantization form
\begin{align}
 \label{eqn:H2nd}
 H_S=\sum_n \epsilon_n a^\dagger_n a_n +\sum_{i,j}^{i\neq j} g_{ij} a^\dagger_i a_j,
\end{align}
where $a_n^\dagger$ and $a_n$ are fermion operators that create and annihilate excitation quanta in state $n$, and $\epsilon_n$ and $g_{ij}$ are site energies and tunnelling constants, respectively.  
The energies and couplings are chosen to correspond to the 
relevant states in the physical system.  We explicitly include the electronic 
ground state $|g\rangle$, which is coupled via the transition dipole $\mu_{ge}$
to a single excitonic state, $|e\rangle$, which in turn is coupled to 
a dark charge transfer state $|ct\rangle$ with electronic 
coupling $g_{\rm xc}$, 
and a single charge separated (polaron) state $|p\rangle$ with 
coupling $g_{\rm xp}$.
We assume that photocurrent arises from the decay of population 
in state $|p\rangle$ into a continuum of bulk polaron states on a timescale much longer 
than any other timescale in our system.  
Fig.\ref{Energy_Feynman_diagrams}a gives schematic of the relative placement of the 
various states explicitly considered in our model. 
We also assume that the internal electronic dynamics are coupled to a dissipative 
environment that modulates the electronic energies as well as electronic transfer integrals
$g_{\rm xc}$ and $g_{\rm xp}$. 

\begin{figure}[tbh]
\centering
\includegraphics[width=\textwidth]{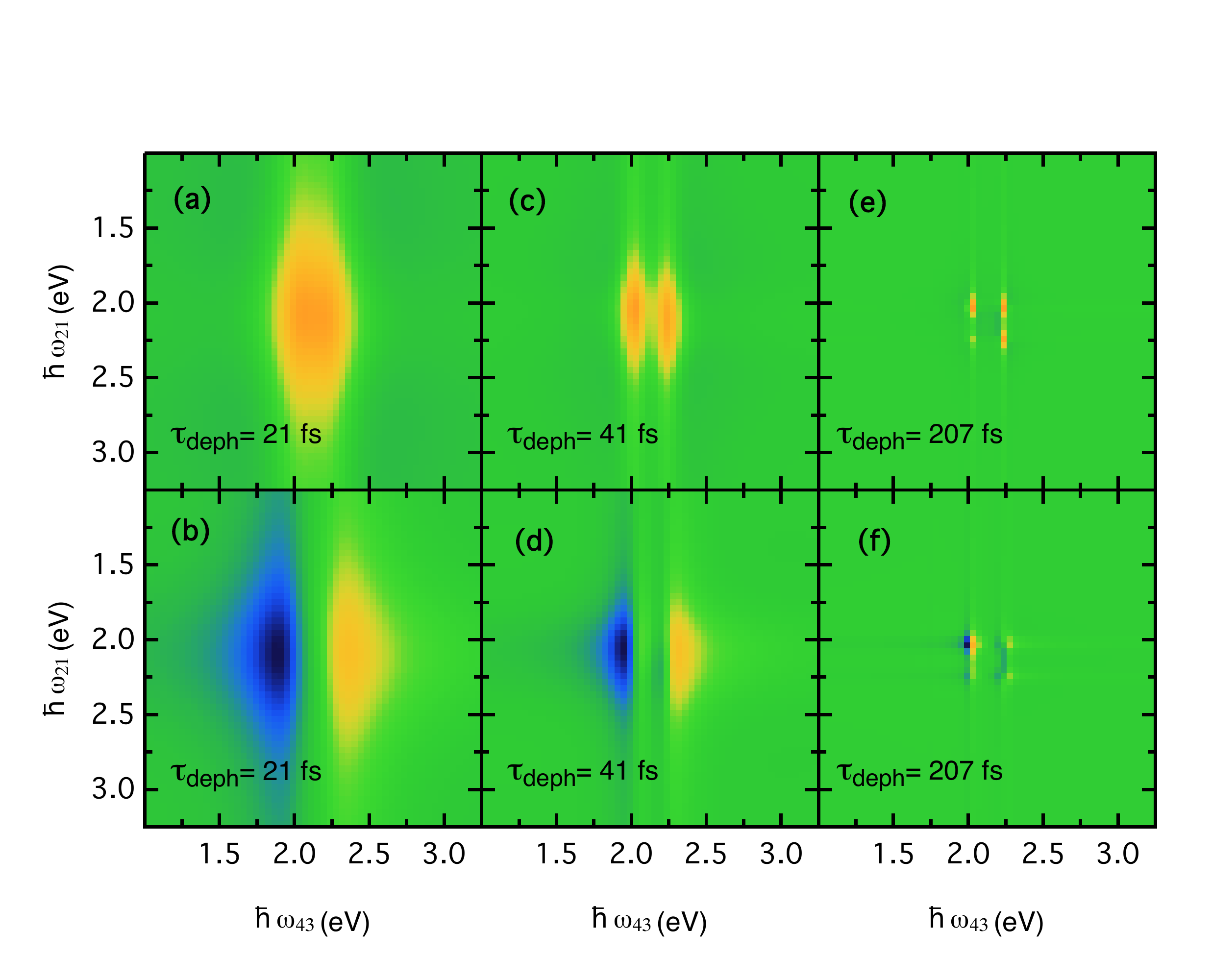}
\caption{Theoretical 2DPCE total correlation functions for the model system with increasing
decoherence times $(\tau_d = \hbar/\gamma)$. The top row shows the real part of the response, while the bottom row shows the imaginary part.}
\label{2D_model}
\end{figure}

We introduce the interactions with the external environment using the Lindblad approach, following a suitable set of assumptions
including that at the initial time, the system and environment are uncorrelated and that the environment is large enough so that its memory time is 
very short. This latter assumption is critical since it allows the use of a Markovian approach. 
The Lindblad approach is commonly used in quantum optics where both of these assumptions are easily met.  For the case at hand, 
in which the environment consists of the electronic polarizibility and vibrational (phonon) response of the molecular surroundings, the 
Markov assumption may be difficult to justify but nonetheless provides a useful starting point for describing the dynamics of our system.
Using Lindblad master equation Eq.~\ref{eqn:LindbladEq}, 
the time evolution of an operator in the system subspace can be determined with the Lindblad dissipation term $\mathscr{L}_D$ given by
\begin{align}
 \label{eqn:Lindbladian}
 \mathscr{L}_D(A)=\sum_k \gamma_k\left[L^\dagger_k AL_k-\frac{1}{2}\left(L^\dagger_k L_k A + A L^\dagger_k L_k \right)\right],
\end{align}
where $\gamma_k$ are the relaxation amplitudes and $L_k$ are the Lindblad operators that characterize the effect of the environment on the system.
Decoherence and relaxation effects on the electronic degrees of freedom 
can be attributed to the oscillation of energy levels 
and the modulation of the electron transfer integral 
via Lindblad operators $L_k=a^\dagger_n a_n$ and $L_k=a^\dagger_i a_j$ $(i\neq j)$. 
Using this approach, we can effectively model both relaxation and decoherence 
within the context of a single theoretical framework, taking the system-environment couplings as 
phenomenological variables.
For example, the timescale for decoherence
 between two electronic states is ultimately set by the fluctuations in the couplings between the 
states {\em viz.}
$$\tau_d^{-1} =\sqrt{\langle\delta V(t)^2\rangle}/\hbar$$ 
where $\delta V(t)$ is a random variable with $\langle \delta V(t)\rangle = 0$ 
stemming from environmental fluctuations ({\em e.g.} 
phonons) described by a single spectral density $S(\omega)$ of the environment.  

In the 2DPCE experiment, it is natural to assume that the total photocurrent signal is proportional to the population of the polaron state $|p \rangle$ 
\begin{align}
 \label{eqn:pc}
  {\cal J} \propto \rho_{\rm pp} = {\rm Tr}[ Q \rho ],
\end{align}
in which $Q=|p\rangle \langle p|$ is the projection operator that 
we take to be the precursor for any photocurrent produced by the pulse sequence. 
We take the light-matter coupling as a weak perturbation and
expand the time-dependent density operator in the powers of the applied laser field as
\begin{align}
 \label{eqn:rho}
  \rho(t)=\rho^{(1)}(t)+\rho^{(2)}(t)+\rho^{(3)}(t)+\cdots.
\end{align}
The polaron population to the $n$th order can be evaluated as $\rho^{(n)}_{\rm pp}(t)={\rm Tr}[Q(t)\rho^{(n)}(t)]$, i.e., 
\begin{align}
 \label{eqn:rhopp}
  \rho_{\rm pp}^{(n)}({\bm r},t) &= \int_0^{\infty}dt_n\int_0^{\infty}dt_{n-1}...\int_0^{\infty}dt_1 S_{\rm po}^{(n)}(t_n,t_{n-1},...,t_1) \nonumber
   \\
   & \times E({\bm r},t-t_n)E({\bm r},t-t_n-t_{n-1})...E({\bm r},t-t_n-t_{n-1}...-t_1),
\end{align}
where $E({\bm r},t)$ is the electric field component of the lasers. 
Note that the time-ordered expansion is in terms of the intervals 
between field-matter interactions denoted by $t_1,t_2,\cdots,t_n$, with $t_1$ being the earliest and $t_n$ being the last. The $n$th order polaron population susceptibility is
\begin{align}
 \label{eqn:snpo}
  S_{\rm po}^{(n)}(t_n,t_{n-1},...,t_1) =\left(\frac{i}{\hbar}\right)^n \langle Q(t_n+...+t_1)[\mu(t_{n-1}+...+t_1),[...[\mu(t_1),[\mu(0),\rho(-\infty)]]...]]\rangle,
\end{align}
where $\mu (t)$ is the electronic dipole operator evolved in time using the Lindblad master equation described above. 
Comparing this to conventional methods for nonlinear spectroscopy, 
such as a four-wave-mixing experiment ending in the ground state, 
the outgoing signal of this photocurrent spectroscopy is governed by the polaron population \cite{AHMarcus:JPCB2012,HSTan:JCP2008}.
It is also important to note that only the even order susceptibilities
are nontrivial according to the electric field symmetry. 
We compute the fourth order susceptibility of the polaron population which is the lowest order to obtain 2D spectrum. 
Consequently, the simplest nonlinear photocurrent spectroscopy corresponds to the 4th order susceptibility.  
A detailed analysis of the susceptibility indicates four distinct contributions 
 that can be categorized as rephasing ($\phi_{43}+\phi_{21}$) and non-rephasing ($\phi_{43}-\phi_{21}$) signals, 
as depicted by the double-side Feynman diagrams in Fig.~\ref{Energy_Feynman_diagrams}, 
corresponding to  
ground-state bleach, stimulated emission, and two excited state absorption pathways. The 
final signal results from interference between each of these possible pathways. 
If, in fact, there is a direct quantum mechanical link between the 
initial excitation and any photocurrent precursor state, 
such links should be revealed by a cross-peak in the 2DPCE signal. However, whether or not 
such cross-peaks can be resolved depends upon the rate of decoherence between the exciton and the precursor state.

In Fig.~\ref{2D_model}
we present the results of our quantum simulations of the 
2DPCE signals produced by our model parametrized to describe the 
PCDTBT:PCBM system. We take the 
decoherence time as a free parameter and compare results at 
increasing coherence times between the 
exciton and the polaron states.  If $\tau_d = 21$\,fs, the 2DPCE response is featureless (Fig.~\ref{2D_model}(a) and (b)).
As the coherence time increases, one should expect to see that cross-peaks begin to 
emerge between the exciton and the polaron states (here at 2.3\,eV).  
Indeed, taking this $\tau_d \gtrsim 30$\,fs 
reveals such cross-peaks (Fig.~\ref{2D_model}(c) and (d)).  In fact, when the decoherence time is taken to be very large (Fig.~\ref{2D_model}(e) and (f)), 
the individual diagonal and off-diagonal spectral components become very well resolved.

\subsection*{2D photocurrent excitation spectroscopy}
\begin{figure}[t]
\centering
\includegraphics[width=0.9\textwidth]{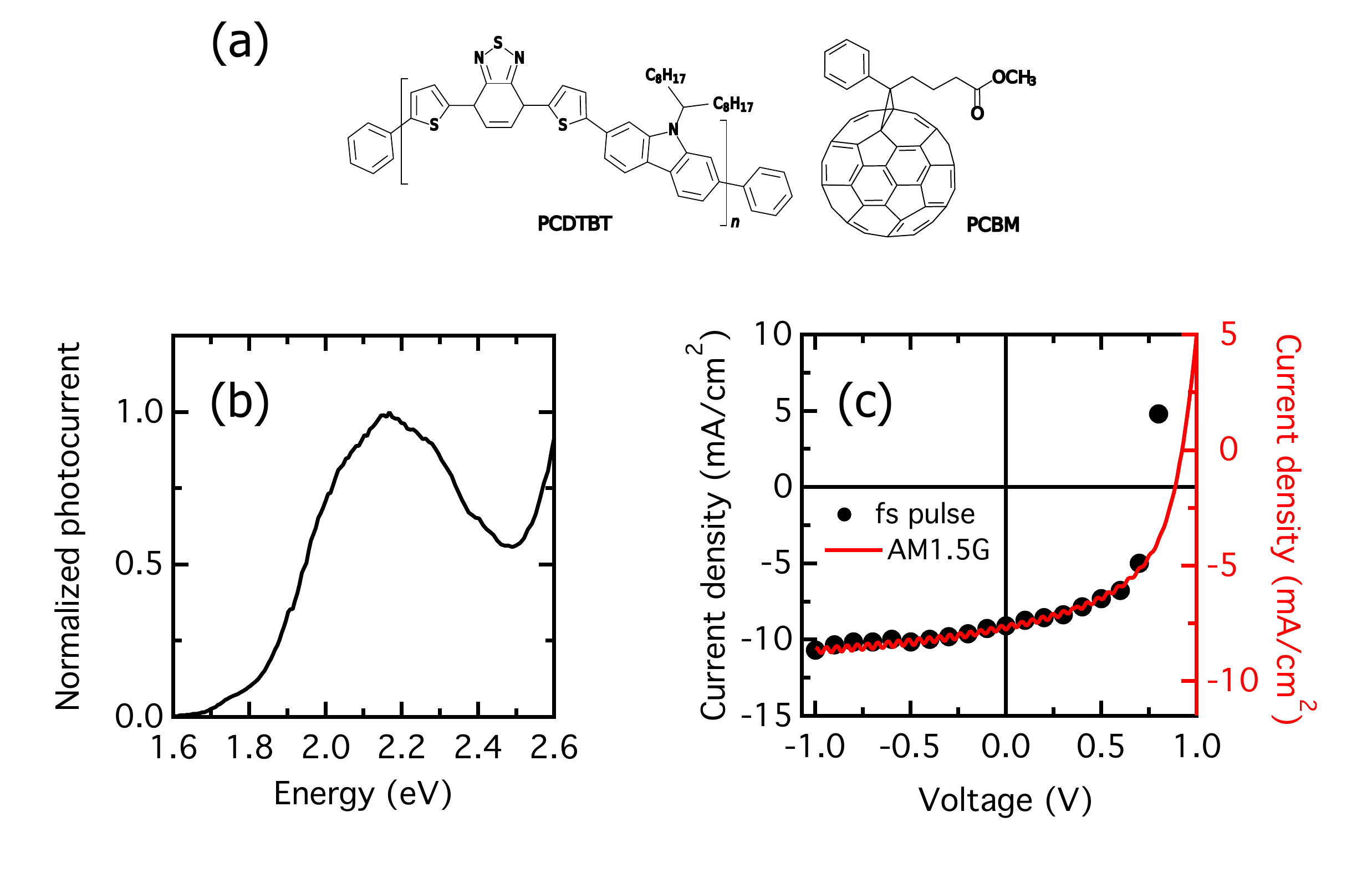}
\caption{(a) Structure of PCDTBT (left) and PCBM (right). (b) Photocurrent excitation spectrum of the PCDTBT:PCBM solar cell under
investigation. 
(c) Current-voltage response of same cell under AM1.5G solar illumination (right-axis) and
under laser illumination (left-axis). }
\label{JV_PLE}
\end{figure}
\begin{figure}[th]
\centering
\includegraphics[width=\textwidth]{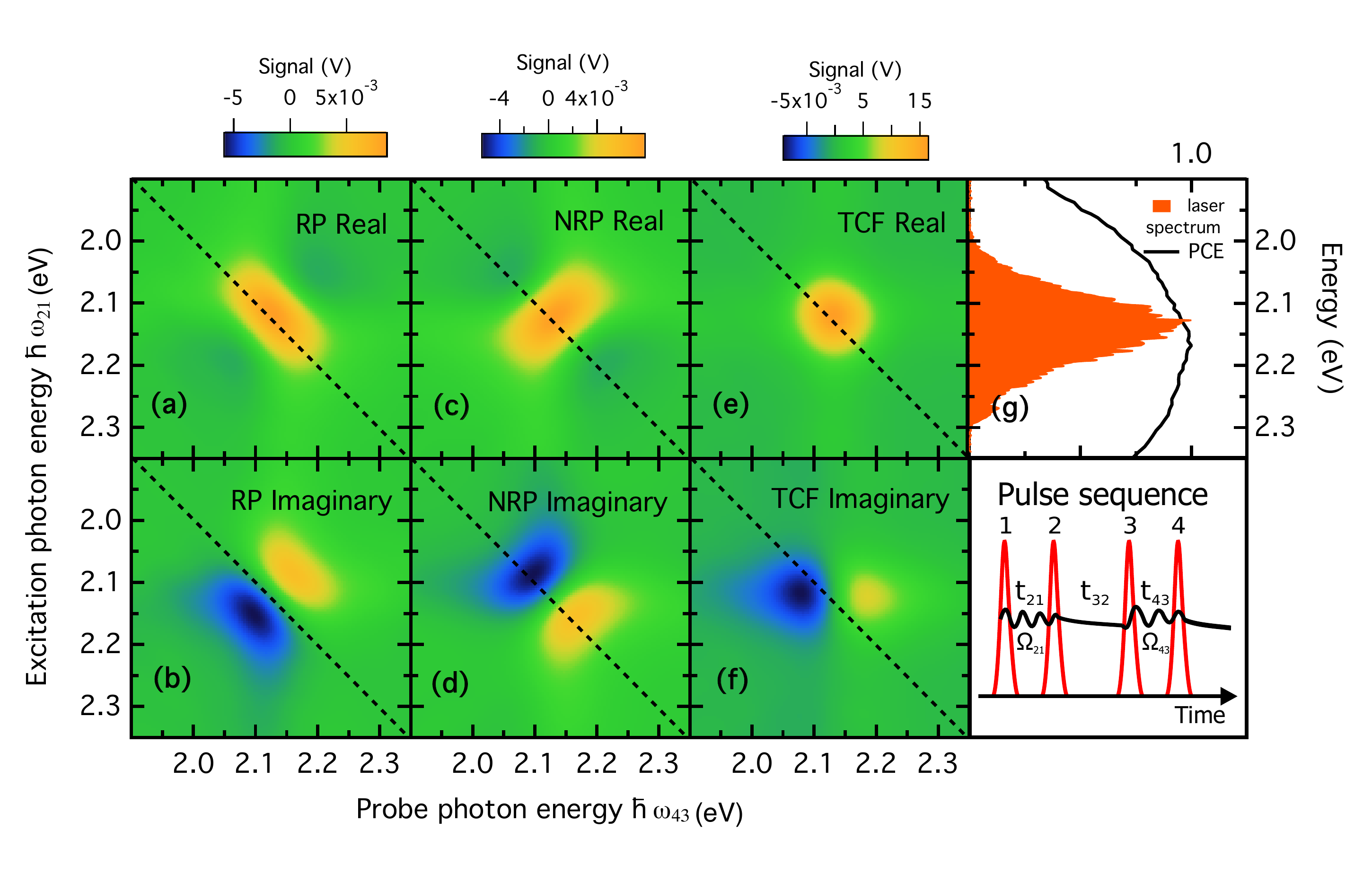}
\caption{Two dimensional coherent photocurrent excitation (2DPCE) spectra at a population waiting time $t_{32}=0$\,fs.
(a,b) Real and imaginary components of the rephasing (RP) signal.
(c,d) Real and imaginary components of the non-rephasing (NRP) signal.
(e,f) Real and imaginary components of the total correlation function (TCF).
(g) Laser spectrum and photocurrent excitation spectra over the same range and schematic of the
experimental pulse sequence.}
\label{2D_experiment}
\end{figure}
We carried out 2DPCE spectroscopy on a solar cell based on PCDTBT:PCBM blend. The molecular structures of PCDTBT and PCBM are shown on Fig.~\ref{JV_PLE}(a) and the device structure is described in the Methods section. This
is a benchmark polymer:fullerene system which has been shown 
to yield solar-power conversion efficiencies as 
high as 7\%.\cite{Park:2009by,Sun:2011si} Fig.~\ref{JV_PLE}(b) presents 
the linear photocurrent excitation spectrum of such solar cell.
2DPCE measurements were performed exciting the solar cell with a 
sequence of four collinear ultrafast laser pulses. Fig.~\ref{JV_PLE}(c) 
reports the current density as a function of external bias of the cell under both 
femtosecond-laser and solar (AM1.5G) illumination. This comparison allows us to conclude that the carrier densities probed by the experiment are comparable to those of the operating solar cell. Working under such photoexcitation densities, 
Fig.~\ref{2D_experiment} presents the 2DPCE spectra of the photovoltaic cell at a population waiting time of $t_{32}=0$\,fs. (see right-most, bottom inset of Fig.~\ref{2D_experiment} for a definition of the inter-pulse delays). Shown are the real and imaginary parts of the rephasing (Fig.~\ref{2D_experiment}(a,b)), nonrephasing (Fig.~\ref{2D_experiment}(c,d)) spectra, and the total correlation function produced by the combination of these (Fig.~\ref{2D_experiment}(e,f)). As evident in Fig.~\ref{2D_experiment}(g), the laser spectrum covers the relevant range for the salient 
electronic states of this system. The total 2DPCE spectra in Fig.~\ref{2D_experiment}(e,f) reveal very little structure of the electronic states of the 
system. The main peak along the diagonal at 2.1\,eV corresponds to the vertical exciton and one expects
to see cross-peaks along the anti-diagonal corresponding to couplings to 
other electronic states coupled to the exciton.  For the case at hand, the other electronic states
of the system are optically dark or at best carry little oscillator strength to the ground state. As shown above by our quantum-dynamical modeling and  Fig.~\ref{2D_model}, 
the fact that we see little trace of 
cross-peaks associated with these states
but do see signal is indicative of very rapid electronic decoherence if resonance tunnelling from excitons to delocalised polarons brought about by bath-induced off-diagonal coupling as suggested in ref.~\citenum{Bittner:2014aa} is the operative mechanism on these early time-scales. 
From these measurements and under the assumptions of our theoretical modelling, we conclude that the decoherence time between exciton and polaron photocarrier precursor states must be $\tau_d \lesssim 20$\,fs.

\begin{figure}[th]
\centering
\includegraphics[width=0.65\textwidth]{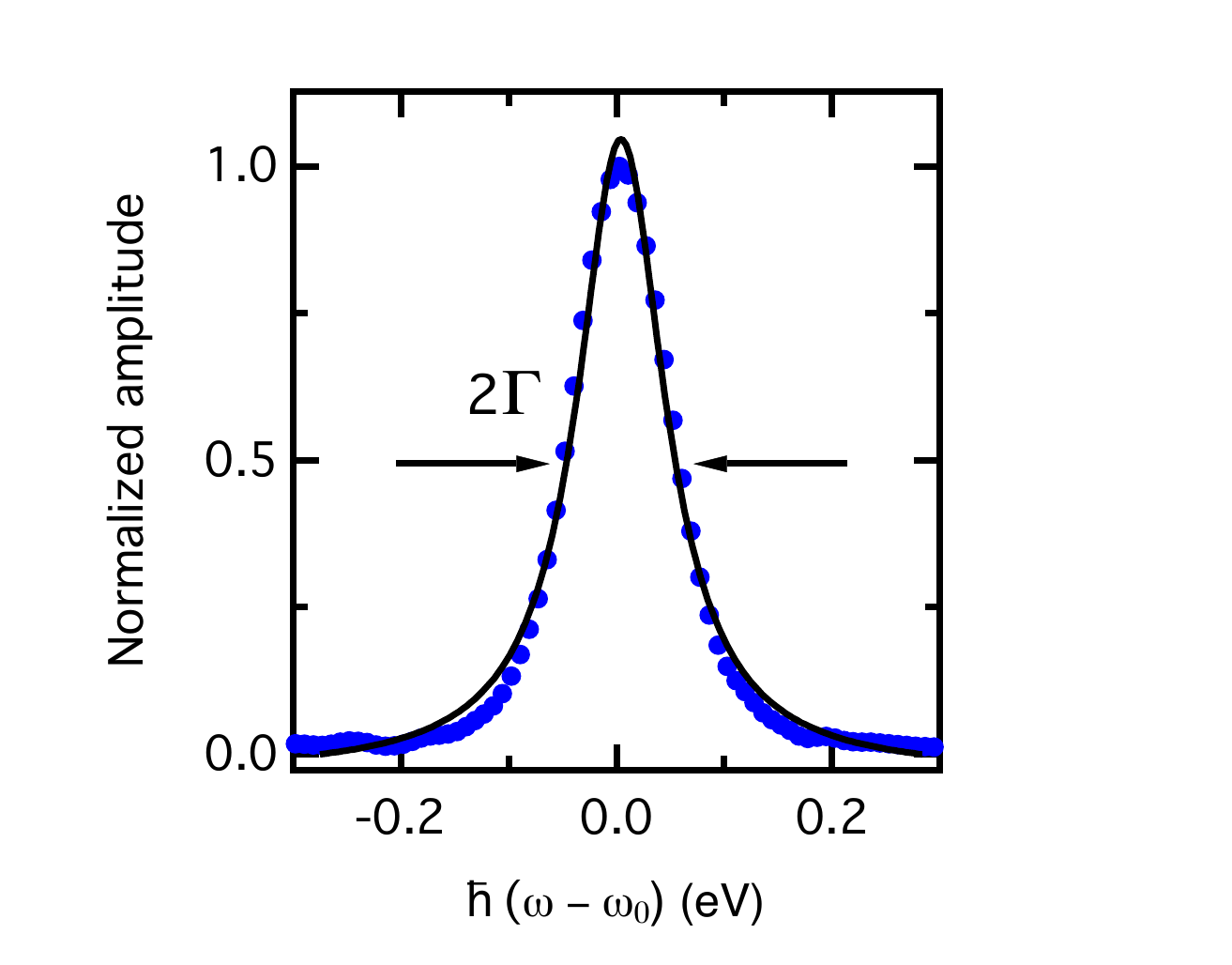}
\caption{Anti-diagonal slice of the 2DPCE rephasing signal. The solid line 
is a fit to a Lorentzian function in which the optical dephasing rate $\Gamma$ characterizes the 
homogeneous linewidth. }
\label{Lorentzian_profile}
\end{figure}
%
In order to explore the decoherence rate inferred above, we seek to measure the optical dephasing time which can be measured from the amplitude of the rephasing spectrum at population waiting time $t_{32} = 0$\,fs (Fig.~\ref{2D_experiment}(a,b)).\cite{Tokmakoff:2000if} Fig.~\ref{Lorentzian_profile} shows such an antidiagonal cut, and the solid line is a fit to a Lorentzian function with full-width-at-half-maximum of $100$\,meV, which represents the homogeneous 
linewidth of the main excitonic transition of PCDTBT over the spectral region covered by the ultrafast laser bandwidth. This implies a pure dephasing time $T_2^* = \hbar / \Gamma = 13$\,fs, where 2$\Gamma$ is the homogeneous linewidth. 
This dephasing time is consistent with the off-diagonal (exciton-polaron) decoherence time inferred above, which we rationalise with the fact that the two quantum dissipation processes are governed by the coupling of the system to the vibronic bath, and are subject to similar spectral density.

\section*{Discussion}
We present here a combined theoretical and experimental investigation into the origins of prompt photocarrier 
production in a working organic photovoltaic system.  The correlation between the observed photocurrent with a sequence of 
ultrashort laser pulses supports our hypothesis that that off-diagonal couplings modulated by 
fluctuations within the bath enable resonant tunneling between an optically prepared exciton 
and a density of delocalized polaron states. 

In a recent paper B\"assler and K\"ohler \cite{Bassler:PCCP2015} reviewed relevant experimental approaches and concepts proposed to describe the charge separation process in donor-acceptor organic solar cells, enforcing the notion that this occurs through a charge separated state that is a vibrationally cold charge transfer state. It is worth noting that our results are not at variance with those conclusions in that our experiments and theoretical modeling aimed to probe the sub-100 fs step of the charge photogeneration, whereas the dissociation process addressed by B\"assler and K\"ohler occurs over time-scales spanning from 100 fs to ns.

The model and experimental results here reported corroborate and tie together 
a number of related recent investigations into the nature of the coupling between excitons and polarons in OPV systems\cite{C5CP05037E,doi:10.1021/ja411859m,doi:10.1021/acs.jpclett.5b00336,Bittner:2014aa}.
For example, in Ref.~\citenum{C5CP05037E}, Kelley and Bittner used mixed quantum/classical methods to sample the 
density of excited states produced by a polymer:fullerene contact pair undergoing molecular dynamics and found that
on a very rapid time-scale, purely excitonic states rapidly mix with charge separated ones even after the 
initial excitation has thermalized with respect to the molecular motions. 
In Ref.~\citenum{doi:10.1021/ja411859m} the authors used {\em ab initio} quantum chemical methods to parameterize a fully quantum model of a polymer:fullerene heterojunction system. 
Quantum dynamics simulations on this system indicate that delocalization of charge in the fullerene phase reduces the Coulomb binding potential 
of the interfacial charge-transfer states and that 
 vibrational excitations provide the necessary energy to 
facilitate charge separation.
Along similar veins, efficient charge mobility can be attributed to the  high degree of packing 
efficiency in the fullerene phase.\cite{doi:10.1021/acs.jpclett.5b00336}
Here, resonant coupling of photogenerated singlet excitons to a high-energy manifold of fullerene electronic states enables efficient charge generation, bypassing localized charge-transfer states. While the theoretical approaches taken by each of these are very different, the underlying picture that emerges 
is excitons decay {\em directly} into states capable of producing photocurrent and that the time-scale for this 
process is ultimately governed by the spectral density of the environment that gives rise to energy level fluctuations. 
These same fluctuations also govern the optical dephasing time of the exciton and set the 
time-scale over which perturbative ({\em i.e.} Fermi golden-rule) descriptions of the decay are applicable.

\section*{Methods}
\subsection*{Solar cell preparation and characterization}
The devices examined in the present work are based on blends of poly(N-90-heptadecanyl-2,7-carbazole-alt-5,5-(40,70- di-2-thienyl-20 ,10 ,30 -benzothiadiazole)) as electron donor and of the fullerene derivative [6,6]-phenyl-C60 butyric acid methyl ester as electron acceptor (PCDTBT:PCBM). 
The PV device structure is ITO/PEDOT:PSS/PCDTBT: PCBM/Ca/Al. PEDOT:PSS was spun onto a cleaned ITO coated glass substrate to form a film of ~35$\,$nm thickness. The active layer of thickness ~80-90$\,$nm was then spin cast on top of PEDOT:PSS layer from a blend of PCDTBT:PCBM (1:2) in chlorobenzene solvent with a concentration of 25mg/ml. The top electrode calcium ($\sim20\,$nm) and aluminium ($\sim100\,$nm) was then subsequently deposited by thermal evaporation. Current density/voltage (J/V) characteristics of the devices were measured using a Keithley 236 Source Measure Unit. Solar cell performance was measured by using a xenon lamp with AM1.5G filters and 100$\,$mW/cm$^{-2}$ illumination  solar simulator (Oriel Instruments).
In the photocurrent excitation measurement, the sample was illuminated at normal incidence by using a modulated laser light as an excitation source, the induced photocurrent was collected by a lock-in amplifier (SR830) referenced to the chopping frequency.

\begin{figure}[tbh]
\centering
\includegraphics[width=\textwidth]{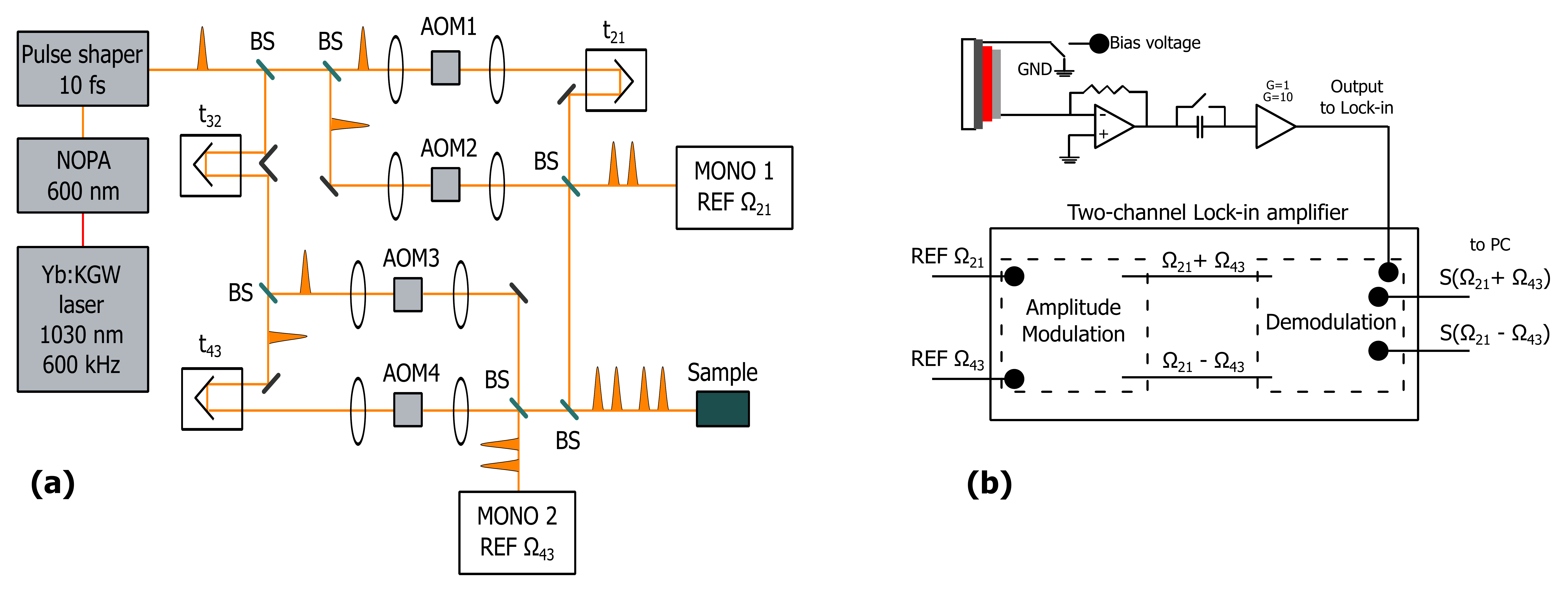}
\caption{(a) Schematic diagram of the 2DPCE setup involving a Mach-Zehnder interferometer with inner Mach-Zehnder interferometers nested in each arm of the outer one. NOPA is noncollinear parametric amplifier, BS is beamsplitter, AOM is acousto-optic modulator, MONO is monochromator, $t_{ij}$ is the temporal delay between pulse $i$ and $j$, and REF is an optically generated reference signal waveform with frequency $\Omega_{ij}$. (b) Schematic diagram of the phase-sensitive detection setup involving amplitude modulation and lockin amplification.}
\label{fig_setup}
\end{figure}

\subsection*{2DPCE Apparatus}
Our 2DPCE setup, shown in Fig.~\ref{fig_setup}, is based on the phase-modulation and phase-sensitive detection system conceived by Marcus and co-workers.\cite{Tekavec:2007rw,Karki:2014eu} It consists of a collinear four-pulse sequence (see bottom-right panel of Fig.~\ref{2D_experiment}) that is generated by means of two Mach-Zehnder interferometers nested in an outer Mach-Zehnder interferometer. The output of a non-collinear optical parametric amplifier (Light Conversion Orpheus-N-3H pumped by a Pharos regenerative amplifier operating at a repetition rate of 600\,kHz) is pre-compressed by an adaptive $4f$ pulse shaper (Biophotonics Solution FemtoJock-P) in order to be $<20$\,fs at the sample position. This output is split into two by means of a 50\% beam splitter, and one arm is delayed with respect to the other by means of an optical delay time defining the delay variable $t_{32}$, and then recombined into a collinear combination of the pulse trains by an identical beam splitter. Each of the two arms  is then split into two sub-arms by identical 50\% beam splitters and then recombined to generate the four-pulse sequence. The twin internal interferometers thus each generate two of the phase-locked pulses of the sequence. In order to ensure the phase locking condition, an acousto-optic Bragg cell is placed in each arm of the twin interferometers. The acousto-optic modulation imposes on each pulse of the sequence a frequency shift equal to the unique modulation frequency $\Omega_i$ ($i$ = 1, 2, 3 and 4), close to 200$\,$MHz. These frequency changes, small compared to the optical frequency, produce a shift in the temporal phase of each pulse which oscillates at the corresponding frequency $\Omega_i$. 
Each laser pulse is phase-shifted with respect to the previous shot, thus giving rise to two collinear trains of phase-modulated pulse pairs. When incident on a photodetector, the interfering excitation pulses produce a population signal oscillating at $\Omega_{21}=\Omega_2-\Omega_1$ and $\Omega_{43}=\Omega_4-\Omega_3$. The exact $\Omega_i$ values are chosen so that $\Omega_{21}$ and $\Omega_{43}$ are in the kHz range (at 5 and 8$\,$kHz respectively, in the present case). This phase modulation scheme allows one to isolate the fourth-order contribution to the photocurrent signal by means of a lockin amplifier (Zurich Instruments HF2LI equipped with multi-frequency and AM/FM modulation modules), which is then used to construct the 2D maps. Reference waveforms of frequency $\Omega_{43}-\Omega_{21}$ and $\Omega_{43}+\Omega_{21}$ are thus generated by amplitude modulation, and are used to demodulate the output of the sample solar cell. A Zurich Instruments HF2TA Current Amplifier is used to convert the current output of the sample device to voltage, as well as to supply an external bias to the device. The non-linear signals of interest oscillate at the frequencies $\Omega_{43}-\Omega_{21}$ and $\Omega_{43}+\Omega_{21}$ and in 2D spectroscopy these are usually referred to respectively as \emph{rephasing} and \emph{non-rephasing} signals. The overall photocurrent signal is filtered by means of a dual lock-in amplifier in order to extract these two frequency components. The 2D spectra are obtained measuring the demodulated signals scanning $t_{21}$ and $t_{43 }$ at fixed $t_{32}$. The three interpulse delay times $t_{21}$, $t_{32}$ and $t_{43}$ are independently computer controlled using three delay stages. The in-phase and the in-quadrature detection channels of the dual lock-in amplifier simultaneously provide the real and imaginary parts of the rephasing and non-rephasing signals for each of such scans. The 2D response function so acquired in the time domain is converted to the energy domain by Fourier-transforming the time variables $t_{21}$ and $t_{43}$ as a parametric function of the $t_{32}$ interpulse time. 
\section*{Acknowledgements}


The work at the University of Houston was funded in part by the
National Science Foundation (CHE-1362006) 
and the Robert A. Welch Foundation (E-1337). 
The work in Montreal was funded by the Natural Sciences and Engineering Research Council of Canada and the University Research Chair in Organic Semiconductor Materials.
The work at Imperial College London was supported by the Engineering and Physical Sciences Research Council via grants EP/K030671, EP/K029843 and EP/J017361 and through a doctoral training studentship (EP/G037515), and by The Royal Society through a Wolfson Merit Award. Montreal and Imperial were supported by an International Exchanges grant from The Royal Society.


\section*{Author contributions statement}

E.B. and C.S. conceived the project.
H.L. developed the model for the spectral simulation calculations and analyzed the results. 
E.V. and P.G. collected and analyzed the experimental 2DPCE data.
S.T. and M.V. fabricated and characterized the devices.
C.B. measured the photocurrent excitation spectrum.
S.F. carried out theoretical {\em ab initio }
calculations on the electronic states of oligomer:PCBM systems.
J.N. directed the device fabrication, 
characterization, and {\em ab initio} calculations. 
E.V. and C.S. also wish to thank Prof. Andrew Marcus for 
many fruitful discussions concerning this work, especially for 
the construction of the 2DPCE apparatus.
All authors contributed equally to this work, but some more 
equally than others.
All authors reviewed the manuscript. 
The authors declare no competing interests.


\begin{thebibliography}{10}
\expandafter\ifx\csname url\endcsname\relax
  \def\url#1{\texttt{#1}}\fi
\expandafter\ifx\csname urlprefix\endcsname\relax\def\urlprefix{URL }\fi
\providecommand{\bibinfo}[2]{#2}
\providecommand{\eprint}[2][]{\url{#2}}

\bibitem{Bredas31012014}
\bibinfo{author}{Bredas, J.-L.}
\newblock \bibinfo{title}{When electrons leave holes in organic solar cells}.
\newblock \emph{\bibinfo{journal}{Science}} \textbf{\bibinfo{volume}{343}},
  \bibinfo{pages}{492--493} (\bibinfo{year}{2014}).
\newblock \urlprefix\url{http://www.sciencemag.org/content/343/6170/492.short}.
\newblock \eprint{http://www.sciencemag.org/content/343/6170/492.full.pdf}.

\bibitem{Gelinas:2011vn}
\bibinfo{author}{G\'elinas, S.} \emph{et~al.}
\newblock \bibinfo{title}{{The Binding Energy of Charge-Transfer Excitons
  Localized at Polymeric Semiconductor Heterojunctions}}.
\newblock \emph{\bibinfo{journal}{J. Phys. Chem. C}}
  \textbf{\bibinfo{volume}{115}}, \bibinfo{pages}{7114--7119}
  (\bibinfo{year}{2011}).
\newblock \urlprefix\url{http://dx.doi.org/10.1021/jp200466y}.

\bibitem{Sariciftci:1994kx}
\bibinfo{author}{Sariciftci, N.} \& \bibinfo{author}{Heeger, A.}
\newblock \bibinfo{title}{Reversible, metastable, ultrafast photoinduced
  electron-transfer from semiconducting polymers to buckminsterfullerene and in
  the corresponding donor-acceptor heterojunctions}.
\newblock \emph{\bibinfo{journal}{International Journal of Modern Physics B}}
  \textbf{\bibinfo{volume}{8}}, \bibinfo{pages}{237--274}
  (\bibinfo{year}{1994}).

\bibitem{Banerji:2010vn}
\bibinfo{author}{Banerji, N.}, \bibinfo{author}{Cowan, S.},
  \bibinfo{author}{Leclerc, M.}, \bibinfo{author}{Vauthey, E.} \&
  \bibinfo{author}{Heeger, A.~J.}
\newblock \bibinfo{title}{{Exciton Formation, Relaxation, and Decay in
  PCDTBT}}.
\newblock \emph{\bibinfo{journal}{{Journal of the American Chemical Society}}}
  \textbf{\bibinfo{volume}{{132}}}, \bibinfo{pages}{17459--17470}
  (\bibinfo{year}{{2010}}).

\bibitem{Tong2010}
\bibinfo{author}{Tong, M.} \emph{et~al.}
\newblock \bibinfo{title}{{Charge carrier photogeneration and decay dynamics in
  the poly(2,7-carbazole) copolymer PCDTBT and in bulk heterojunction
  composites with PC$_{70}$BM}}.
\newblock \emph{\bibinfo{journal}{Phys. Rev. B}} \textbf{\bibinfo{volume}{81}},
  \bibinfo{pages}{125210} (\bibinfo{year}{2010}).

\bibitem{Sheng2012}
\bibinfo{author}{Sheng, C.-X.}, \bibinfo{author}{Basel, T.},
  \bibinfo{author}{Pandit, B.} \& \bibinfo{author}{Vardeny, Z.~V.}
\newblock \bibinfo{title}{{Photoexcitation dynamics in polythiophene/fullerene
  blends for photovoltaic applications}}.
\newblock \emph{\bibinfo{journal}{Organic Electronics}}
  \textbf{\bibinfo{volume}{13}}, \bibinfo{pages}{1031--1037}
  (\bibinfo{year}{2012}).

\bibitem{Jailaubekov:2013fk}
\bibinfo{author}{Jailaubekov, A.~E.} \emph{et~al.}
\newblock \bibinfo{title}{Hot charge-transfer excitons set the time limit for
  charge separation at donor/acceptor interfaces in organic photovoltaics}.
\newblock \emph{\bibinfo{journal}{Nat Mater}} \textbf{\bibinfo{volume}{12}},
  \bibinfo{pages}{66--73} (\bibinfo{year}{2013}).
\newblock \urlprefix\url{http://dx.doi.org/10.1038/nmat3500}.

\bibitem{Grancini:2013uq}
\bibinfo{author}{Grancini, G.} \emph{et~al.}
\newblock \bibinfo{title}{Hot exciton dissociation in polymer solar cells}.
\newblock \emph{\bibinfo{journal}{Nat Mater}} \textbf{\bibinfo{volume}{12}},
  \bibinfo{pages}{29--33} (\bibinfo{year}{2013}).
\newblock \urlprefix\url{http://dx.doi.org/10.1038/nmat3502}.

\bibitem{doi:10.1021/jz4010569}
\bibinfo{author}{Kaake, L.~G.}, \bibinfo{author}{Moses, D.} \&
  \bibinfo{author}{Heeger, A.~J.}
\newblock \bibinfo{title}{Coherence and uncertainty in nanostructured organic
  photovoltaics}.
\newblock \emph{\bibinfo{journal}{The Journal of Physical Chemistry Letters}}
  \textbf{\bibinfo{volume}{4}}, \bibinfo{pages}{2264--2268}
  (\bibinfo{year}{2013}).
\newblock \urlprefix\url{http://pubs.acs.org/doi/abs/10.1021/jz4010569}.
\newblock \eprint{http://pubs.acs.org/doi/pdf/10.1021/jz4010569}.

\bibitem{doi:10.1021/jp4071086}
\bibinfo{author}{Mukamel, S.}
\newblock \bibinfo{title}{Comment on ``{C}oherence and uncertainty in
  nanostructured organic photovoltaics"}.
\newblock \emph{\bibinfo{journal}{The Journal of Physical Chemistry A}}
  \textbf{\bibinfo{volume}{117}}, \bibinfo{pages}{10563--10564}
  (\bibinfo{year}{2013}).
\newblock \urlprefix\url{http://pubs.acs.org/doi/abs/10.1021/jp4071086}.
\newblock \eprint{http://pubs.acs.org/doi/pdf/10.1021/jp4071086}.

\bibitem{Banerji:2013ej}
\bibinfo{author}{Banerji, N.}
\newblock \bibinfo{title}{{Sub-picosecond delocalization in the excited state
  of conjugated homopolymers and donor--acceptor copolymers}}.
\newblock \emph{\bibinfo{journal}{J. Mater. Chem. C}}
  \textbf{\bibinfo{volume}{1}}, \bibinfo{pages}{3052--3066}
  (\bibinfo{year}{2013}).

\bibitem{Gelinas:2013fk}
\bibinfo{author}{G{\'e}linas, S.} \emph{et~al.}
\newblock \bibinfo{title}{Coherent charge separation in organic semiconductor
  photovoltaic diodes}.
\newblock \emph{\bibinfo{journal}{Science}} \textbf{\bibinfo{volume}{343}},
  \bibinfo{pages}{521--516} (\bibinfo{year}{2014}).

\bibitem{Provencher:fk}
\bibinfo{author}{Provencher, F.} \emph{et~al.}
\newblock \bibinfo{title}{Direct observation of ultrafast long-range charge
  separation at polymer:fullerene heterojunctions}.
\newblock \emph{\bibinfo{journal}{Nat Commun}} \textbf{\bibinfo{volume}{5}},
  \bibinfo{pages}{4288} (\bibinfo{year}{2014}).

\bibitem{Barbara:1996uc}
\bibinfo{author}{Barbara, P.}, \bibinfo{author}{Meyer, T.} \&
  \bibinfo{author}{Ratner, M.}
\newblock \bibinfo{title}{{Contemporary issues in electron transfer research}}.
\newblock \emph{\bibinfo{journal}{J Phys Chem-Us}}
  \textbf{\bibinfo{volume}{100}}, \bibinfo{pages}{13148--13168}
  (\bibinfo{year}{1996}).

\bibitem{Falke30052014}
\bibinfo{author}{Falke, S.~M.} \emph{et~al.}
\newblock \bibinfo{title}{Coherent ultrafast charge transfer in an organic
  photovoltaic blend}.
\newblock \emph{\bibinfo{journal}{Science}} \textbf{\bibinfo{volume}{344}},
  \bibinfo{pages}{1001--1005} (\bibinfo{year}{2014}).
\newblock
  \urlprefix\url{http://www.sciencemag.org/content/344/6187/1001.abstract}.
\newblock \eprint{http://www.sciencemag.org/content/344/6187/1001.full.pdf}.

\bibitem{Ishizaki:2009gd}
\bibinfo{author}{Ishizaki, A.} \& \bibinfo{author}{Fleming, G.~R.}
\newblock \bibinfo{title}{Theoretical examination of quantum coherence in a
  photosynthetic system at physiological temperature}.
\newblock \emph{\bibinfo{journal}{Proceedings of the National Academy of
  Sciences}} \textbf{\bibinfo{volume}{106}}, \bibinfo{pages}{17255--17260}
  (\bibinfo{year}{2009}).
\newblock \urlprefix\url{http://www.pnas.org/content/106/41/17255.abstract}.

\bibitem{Song:1uq}
\bibinfo{author}{Song, Y.}, \bibinfo{author}{Clafton, S.~N.},
  \bibinfo{author}{Pensack, R.~D.}, \bibinfo{author}{Kee, T.~W.} \&
  \bibinfo{author}{Scholes, G.~D.}
\newblock \bibinfo{title}{{Vibrational coherence probes the mechanism of
  ultrafast electron transfer in polymer{\&}ndash;fullerene blends}}.
\newblock \emph{\bibinfo{journal}{Nature Communications}}
  \textbf{\bibinfo{volume}{5}}, \bibinfo{pages}{4933} (\bibinfo{year}{2014}).
\newblock
  \urlprefix\url{http://dx.doi.org/10.1038/ncomms5933$\backslash$npapers3://publication/doi/10.1038/ncomms5933}.
\newblock \eprint{0402594v3}.

\bibitem{Rozzi:2013fk}
\bibinfo{author}{Andrea~Rozzi, C.} \emph{et~al.}
\newblock \bibinfo{title}{Quantum coherence controls the charge separation in a
  prototypical artificial light-harvesting system}.
\newblock \emph{\bibinfo{journal}{Nat Commun}} \textbf{\bibinfo{volume}{4}},
  \bibinfo{pages}{1602} (\bibinfo{year}{2013}).
\newblock \urlprefix\url{http://dx.doi.org/10.1038/ncomms2603}.

\bibitem{Bittner:2014aa}
\bibinfo{author}{Bittner, E.~R.} \& \bibinfo{author}{Silva, C.}
\newblock \bibinfo{title}{Noise-induced quantum coherence drives photo-carrier
  generation dynamics at polymeric semiconductor heterojunctions}.
\newblock \emph{\bibinfo{journal}{Nat Commun}} \textbf{\bibinfo{volume}{5}},
  \bibinfo{pages}{3119} (\bibinfo{year}{2014}).
\newblock \urlprefix\url{http://dx.doi.org/10.1038/ncomms4119}.

\bibitem{Yao:2015aa}
\bibinfo{author}{Yao, Y.}, \bibinfo{author}{Zhou, N.}, \bibinfo{author}{Prior,
  J.} \& \bibinfo{author}{Zhao, Y.}
\newblock \bibinfo{title}{Competition between diagonal and off-diagonal
  coupling gives rise to charge-transfer states in polymeric solar cells}.
\newblock \emph{\bibinfo{journal}{Scientific Reports}}
  \textbf{\bibinfo{volume}{5}}, \bibinfo{pages}{14555} (\bibinfo{year}{2015}).
\newblock \urlprefix\url{http://www.ncbi.nlm.nih.gov/pmc/articles/PMC4585960/}.

\bibitem{Zhao-2014}
\bibinfo{author}{Zhao, Y.}, \bibinfo{author}{Yao, Y.},
  \bibinfo{author}{Chernyak, V.} \& \bibinfo{author}{Zhao, Y.}
\newblock \bibinfo{title}{Communication: Spin-boson model with diagonal and
  off-diagonal coupling to two independent baths: Ground-state phase transition
  in the deep sub-ohmic regime}.
\newblock \emph{\bibinfo{journal}{J Chem Phys}} \textbf{\bibinfo{volume}{140}},
  \bibinfo{pages}{161105} (\bibinfo{year}{2014}).

\bibitem{Bakulin:2016rt}
\bibinfo{author}{Bakulin, A.~A.}, \bibinfo{author}{Silva, C.} \&
  \bibinfo{author}{Vella, E.}
\newblock \bibinfo{title}{Ultrafast spectroscopy with photocurrent detection:
  Watching excitonic optoelectronic systems at work}.
\newblock \emph{\bibinfo{journal}{J. Phys. Chem. Lett.}}
  \textbf{\bibinfo{volume}{7}}, \bibinfo{pages}{250--258}
  (\bibinfo{year}{2016}).
\newblock \urlprefix\url{http://dx.doi.org/10.1021/acs.jpclett.5b01955}.

\bibitem{Nardin:2013bf}
\bibinfo{author}{Nardin, G.}, \bibinfo{author}{Autry, T.~M.},
  \bibinfo{author}{Silverman, K.~L.} \& \bibinfo{author}{Cundiff, S.~T.}
\newblock \bibinfo{title}{Multidimensional coherent photocurrent spectroscopy
  of a semiconductor nanostructure}.
\newblock \emph{\bibinfo{journal}{Opt. Express}} \textbf{\bibinfo{volume}{21}},
  \bibinfo{pages}{28617--28627} (\bibinfo{year}{2013}).
\newblock \urlprefix\url{http://dx.doi.org/10.1364/OE.21.028617}.

\bibitem{Karki:2014eu}
\bibinfo{author}{Karki, K.~J.} \emph{et~al.}
\newblock \bibinfo{title}{{Coherent two-dimensional photocurrent spectroscopy
  in a PbS quantum dot photocell}}.
\newblock \emph{\bibinfo{journal}{Nat. Commun.}} \textbf{\bibinfo{volume}{5}},
  \bibinfo{pages}{5869} (\bibinfo{year}{2014}).
\newblock \urlprefix\url{http://dx.doi.org/10.1038/ncomms6869}.

\bibitem{Few:2015fk}
\bibinfo{author}{Few, S.}, \bibinfo{author}{Frost, J.~M.} \&
  \bibinfo{author}{Nelson, J.}
\newblock \bibinfo{title}{{Models of charge pair generation in organic solar
  cells}}.
\newblock \emph{\bibinfo{journal}{Phys. Chem. Chem. Phys.}}
  \textbf{\bibinfo{volume}{17}}, \bibinfo{pages}{2311--2325}
  (\bibinfo{year}{2015}).
\newblock \urlprefix\url{http://xlink.rsc.org/?DOI=C4CP03663H}.
\newblock \eprint{{\_}barata Materials and Techniques of polychrome wooden
  sculpture}.

\bibitem{Few:2014fk}
\bibinfo{author}{Few, S.}, \bibinfo{author}{Frost, J.~M.},
  \bibinfo{author}{Kirkpatrick, J.} \& \bibinfo{author}{Nelson, J.}
\newblock \bibinfo{title}{{Influence of chemical structure on the charge
  transfer state spectrum of a polymer:fullerene complex}}.
\newblock \emph{\bibinfo{journal}{Journal of Physical Chemistry C}}
  \textbf{\bibinfo{volume}{118}}, \bibinfo{pages}{8253--8261}
  (\bibinfo{year}{2014}).

\bibitem{bakulin2012}
\bibinfo{author}{Bakulin, A.~A.} \emph{et~al.}
\newblock \bibinfo{title}{The role of driving energy and delocalized states for
  charge separation in organic semiconductors}.
\newblock \emph{\bibinfo{journal}{Science}} \textbf{\bibinfo{volume}{335}},
  \bibinfo{pages}{1340--1344} (\bibinfo{year}{2012}).
\newblock \urlprefix\url{http://science.sciencemag.org/content/335/6074/1340}.
\newblock
  \eprint{http://science.sciencemag.org/content/335/6074/1340.full.pdf}.

\bibitem{AHMarcus:JPCB2012}
\bibinfo{author}{Perdomo-Ortiz, A.}, \bibinfo{author}{Widom, J.~R.},
  \bibinfo{author}{Lott, G.~A.}, \bibinfo{author}{Aspuru-Guzik, A.} \&
  \bibinfo{author}{Marcus, A.~H.}
\newblock \bibinfo{title}{Conformation and electronic population transfer in
  membrane-supported self-assembled porphyrin dimers by 2d fluorescence
  spectroscopy}.
\newblock \emph{\bibinfo{journal}{J. Phys. Chem. B}}
  \textbf{\bibinfo{volume}{116}}, \bibinfo{pages}{10757--10770}
  (\bibinfo{year}{2012}).
\newblock \urlprefix\url{http://pubs.acs.org/doi/abs/10.1021/jp305916x}.

\bibitem{HSTan:JCP2008}
\bibinfo{author}{Tan, H.-S.}
\newblock \bibinfo{title}{Theory and phase-cycling scheme selection principles
  of collinear phase coherent multi-dimensional optical spectroscopy}.
\newblock \emph{\bibinfo{journal}{The Journal of Chemical Physics}}
  \textbf{\bibinfo{volume}{129}} (\bibinfo{year}{2008}).
\newblock
  \urlprefix\url{http://scitation.aip.org/content/aip/journal/jcp/129/12/10.1063/1.2978381}.

\bibitem{Park:2009by}
\bibinfo{author}{Park, S.~H.} \emph{et~al.}
\newblock \bibinfo{title}{{Bulk heterojunction solar cells with internal
  quantum efficiency approaching 100{\%}}}.
\newblock \emph{\bibinfo{journal}{Nat. Photonics}}
  \textbf{\bibinfo{volume}{3}}, \bibinfo{pages}{297--302}
  (\bibinfo{year}{2009}).
\newblock \urlprefix\url{http://dx.doi.org/10.1038/NPHOTON.2009.69}.

\bibitem{Sun:2011si}
\bibinfo{author}{Sun, Y.} \emph{et~al.}
\newblock \bibinfo{title}{{Efficient, Air-Stable Bulk Heterojunction Polymer
  Solar Cells Using MoOx as the Anode Interfacial Layer}}.
\newblock \emph{\bibinfo{journal}{Advanced Materials}}
  \textbf{\bibinfo{volume}{23}}, \bibinfo{pages}{2226--2230}
  (\bibinfo{year}{2011}).
\newblock \urlprefix\url{http://dx.doi.org/10.1002/adma.201100038}.

\bibitem{Tokmakoff:2000if}
\bibinfo{author}{Tokmakoff, A.}
\newblock \bibinfo{title}{{Two-Dimensional Line Shapes Derived from Coherent
  Third-Order Nonlinear Spectroscopy}}.
\newblock \emph{\bibinfo{journal}{J. Phys. Chem. A}}
  \textbf{\bibinfo{volume}{104}}, \bibinfo{pages}{4247--4255}
  (\bibinfo{year}{2000}).
\newblock \urlprefix\url{http://dx.doi.org/10.1021/jp993207r}.

\bibitem{Bassler:PCCP2015}
\bibinfo{author}{B\"assler, H.} \& \bibinfo{author}{K\"ohler, A.}
\newblock \bibinfo{title}{{"}hot or cold{"}: how do charge transfer states at
  the donor-acceptor interface of an organic solar cell dissociate?}
\newblock \emph{\bibinfo{journal}{Phys. Chem. Chem. Phys.}}
  \textbf{\bibinfo{volume}{17}}, \bibinfo{pages}{28451--28462}
  (\bibinfo{year}{2015}).
\newblock \urlprefix\url{http://dx.doi.org/10.1039/C5CP04110D}.

\bibitem{C5CP05037E}
\bibinfo{author}{Bittner, E.~R.} \& \bibinfo{author}{Kelley, A.}
\newblock \bibinfo{title}{The role of structural fluctuations and environmental
  noise in the electron/hole separation kinetics at organic polymer
  bulk-heterojunction interfaces}.
\newblock \emph{\bibinfo{journal}{Phys. Chem. Chem. Phys.}}
  \textbf{\bibinfo{volume}{17}}, \bibinfo{pages}{28853--28859}
  (\bibinfo{year}{2015}).
\newblock \urlprefix\url{http://dx.doi.org/10.1039/C5CP05037E}.

\bibitem{doi:10.1021/ja411859m}
\bibinfo{author}{Savoie, B.~M.} \emph{et~al.}
\newblock \bibinfo{title}{Unequal partnership: Asymmetric roles of polymeric
  donor and fullerene acceptor in generating free charge}.
\newblock \emph{\bibinfo{journal}{Journal of the American Chemical Society}}
  \textbf{\bibinfo{volume}{136}}, \bibinfo{pages}{2876--2884}
  (\bibinfo{year}{2014}).
\newblock \urlprefix\url{http://dx.doi.org/10.1021/ja411859m}.
\newblock \bibinfo{note}{PMID: 24460057},
  \eprint{http://dx.doi.org/10.1021/ja411859m}.

\bibitem{doi:10.1021/acs.jpclett.5b00336}
\bibinfo{author}{Huix-Rotllant, M.}, \bibinfo{author}{Tamura, H.} \&
  \bibinfo{author}{Burghardt, I.}
\newblock \bibinfo{title}{Concurrent effects of delocalization and internal
  conversion tune charge separation at regioregular polythiophene--fullerene
  heterojunctions}.
\newblock \emph{\bibinfo{journal}{The Journal of Physical Chemistry Letters}}
  \textbf{\bibinfo{volume}{6}}, \bibinfo{pages}{1702--1708}
  (\bibinfo{year}{2015}).
\newblock \urlprefix\url{http://dx.doi.org/10.1021/acs.jpclett.5b00336}.
\newblock \bibinfo{note}{PMID: 26263337},
  \eprint{http://dx.doi.org/10.1021/acs.jpclett.5b00336}.

\bibitem{Tekavec:2007rw}
\bibinfo{author}{Tekavec, P.~F.}, \bibinfo{author}{Lott, G.~A.} \&
  \bibinfo{author}{Marcus, A.~H.}
\newblock \bibinfo{title}{{Fluorescence-detected two-dimensional electronic
  coherence spectroscopy by acousto-optic phase modulation}}.
\newblock \emph{\bibinfo{journal}{J. Chem. Phys.}}
  \textbf{\bibinfo{volume}{127}}, \bibinfo{pages}{214307}
  (\bibinfo{year}{2007}).
\newblock \urlprefix\url{http://dx.doi.org/10.1063/1.2800560}.

\end{thebibliography}

\end{document}